\documentclass[11pt,letterpaper]{article}
\usepackage{emnlp2017}
\usepackage{times}
\usepackage{latexsym}

\usepackage{graphicx}
\usepackage{float}
\usepackage{todonotes}

\emnlpfinalcopy

\def\emnlppaperid{***}

\title{TableQA: Question Answering on Tabular Data}

\author{Svitlana Vakulenko \and Vadim Savenkov \\
  {\tt svitlana.vakulenko@wu.ac.at}\\
  {\tt vadim.savenkov@wu.ac.at}}

\date{}

\begin{document}

\maketitle


\begin{abstract}
Tabular data is difficult to analyze and to search through, yielding for new tools and interfaces that would allow even non tech-savvy users to gain insights from open datasets without resorting to specialized data analysis tools or even without having to fully understand the dataset structure. The goal of our demonstration is to showcase answering natural language questions from tabular data, and to discuss related system configuration and model training aspects. Our prototype is publicly available and open-sourced (see \url{https://svakulenko.ai.wu.ac.at/tableqa}).
\end{abstract}

\section{Introduction}

There is an abundance of tabular data on the web in the form of Open Data tables, which are regularly released by many national governments. Providing their data free of charge, publishing bodies seldom have dedicated resources to support the end users in finding and using it. In many open data portals the search facility remains limited: e.g., no search in the content of data tables is supported. 

We attempt to remedy this situation through development of the information retrieval tools tailored specifically to the end users without technical background. Our Open Data Assistant chatbot~\cite{us} offers an unconventional interface for cross-lingual data search via Facebook and Skype messaging applications enabling a quick overview of the available datasets collected from various open data portals. However, the current version of the chatbot supports only metadata-based search. In this paper, we work towards extending the chatbot to search within the content of open data tables and answering specific user questions using the values from these tables.

\begin{figure*}[t!]
    \centering
    \includegraphics[width=0.8\textwidth]{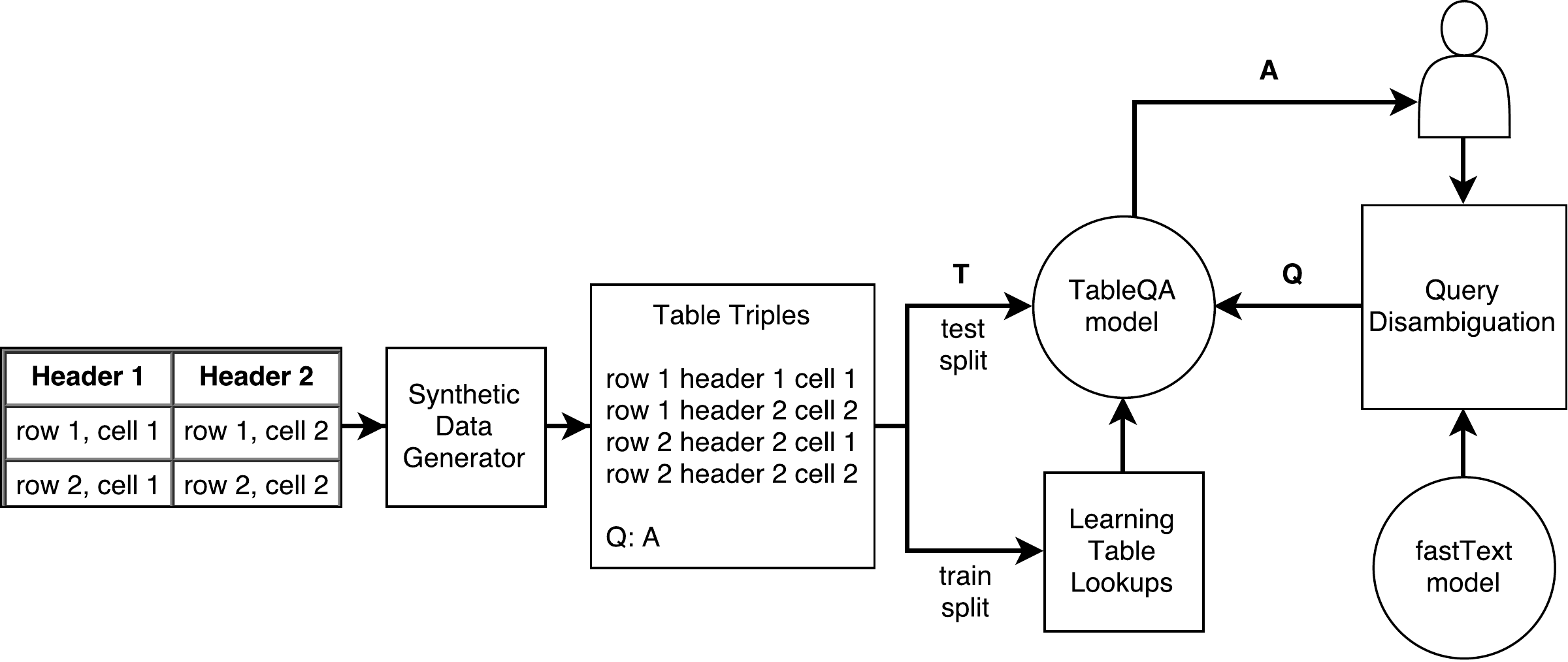}
    \caption{System architecture. T - input table; Q - question; A - answer}
    \label{fig:arch}
\end{figure*}

\section{Task Description}

The task of question answering over tables is given an input table (or a set of tables) \textit{T} and a natural language question \textit{Q} (a user query), output the correct answer \textit{A}.

\section{Related Work}

Recently, quite a few studies emerged that address the question-answering task on tables using deep neural networks. They involve search across tables~\cite{DBLP:conf/www/SunMHYSY16} and learning to perform aggregation operations~\cite{DBLP:conf/ijcai/YinLLK16,DBLP:journals/corr/NeelakantanLAMA16}. However, all of the proposed systems are very complex, require significant computation resources and are engineered to work exclusively on tabular data.

We contribute to the growing body of work on question answering for tabular data by providing and evaluating a prototype based on the End-To-End Memory Networks architecture~\cite{DBLP:conf/nips/SukhbaatarSWF15}. This architecture was originally designed for the question-answering tasks from short natural language texts (bAbI tasks)~\cite{DBLP:journals/corr/WestonBCM15}, which include testing elements of inductive and deductive reasoning, co-reference resolution and time manipulation. In this context the task of question answering over tables can be seen as an extension to the original bAbI tasks. It is very appealing to be able to apply the same type of architecture to querying semi-structured tables alongside the textual data for this could enable question answering on real-world documents that contain a mixture of both, e.g., user manuals and financial reports.

\section{Architecture}

The architecture of our system for table-based question answering is summarized in Figure~\ref{fig:arch}. Each of the individual components is described in further details below.

\subsection{Table Representation}

Training examples consist of the input table decomposed into row-column-value triples and a question/answer pair, for instance: 

\begin{tabular}{lll}
\hline
Row1 & City & Klagenfurt\\
Row1 & Immigration & 110\\
Row1 & Emmigration & 140\\
Row2 & City & Salzburg\\
Row2 & Immigration & 170\\
Row2 & Emmigration & 100\\
\hline
\end{tabular}\\[1ex]
\textbf{Question}: What is the immigration in Salzburg?\\
\textbf{Answer}: 170

This representation preserves the row and column identifiers of the table values. In this way our system can also ingest and learn from multiple tables at once. 

\subsection{Learning Table Lookups}

Our method for question answering from tables is based on the End-To-End Memory Network architecture~\cite{DBLP:conf/nips/SukhbaatarSWF15}, which we employ to transform the natural-language questions into the table lookups. Memory Network is a recurrent neural network (RNN) trained to predict the correct answer by combining continuous representations of an input table and a question. It consists of a sequence of memory layers (3 layers in our experiments) that allow to go over the content of the input table several times and perform reasoning in multiple steps.

The data samples for training and testing are fed in batches (batch size is 32 in our experiments). Each of the data samples consists of the input table, a question and the correct answer that corresponds to one of the cells in the input table. 


The input tables, questions and answers are embedded into a vector space using a bag-of-words models, which neglects the ordering of words. The output layer generates the predicted answer to the input question and is implemented as a softmax function in the size of the vocabulary, i.e. it outputs the probability distribution over all possible answers, which could be any of the table cells.

The network is trained using stochastic gradient descent with linear start to avoid the local minima as in~\cite{DBLP:conf/nips/SukhbaatarSWF15}. The objective function is to minimize the cross-entropy loss between the predicted answer and the true answer from the training set.

\begin{figure*}[t!]
    \centering
    \includegraphics[width=0.9\textwidth]{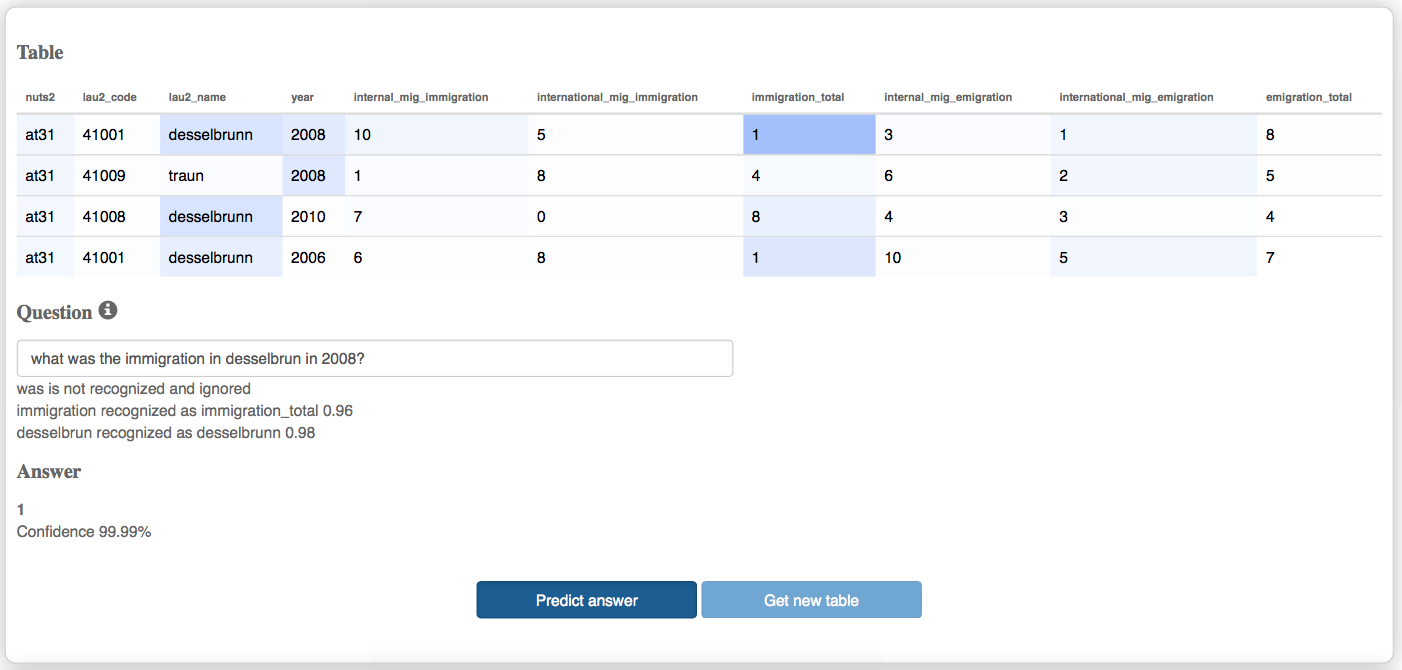}
    \caption{Demonstration of the neural network model trained towards question answering from tables. The highlighted cells and their intensity indicate the attention weights over the input table that were activated to predict the answer to the input question.}
    \label{fig:ui}
\end{figure*}

\subsection{Query Disambiguation}

Since users may refer to the columns with words that differ from the labels used in the table headings, we employ a fastText model~\cite{bojanowski2016enriching} pretrained on Wikipedia to compute similarity between the out-of-vocabulary (OOV) words from the user query and the words in our vocabulary, i.e. to align or ground the query in the local representation. 

fastText provides continuous word representation, which reflects semantic similarity using both the word co-occurrence statistics and the sub-word-based similarity via the character n-grams. For each of the OOV words the query disambiguation module picks the most similar word from the vocabulary at query time and uses its embedding instead.

In our scenario this approach is particularly useful to match the paraphrases of the column headings, e.g., the word \textit{emigration} is matched to the \textit{emigration\_total} label. We empirically learned the similarity threshold of 0.8 that provides optimal precision/recall trade-off on our data.

\section{Experiments}
\subsection{Synthetic data}

We produce synthetic training examples based on a real-world table by limiting the domain for each of the column-variables to n distinct values per column (10 in our experiments). From this vocabulary we generate sample tables and question/answer pairs using the predefined templates. Thus, the vocabulary size in our experiments was fixed to 65 words.

For each table we generate 4 unique rows and a question addressed towards a cell from one of these rows. The templates produce two types of questions that model functional dependencies in the table with

(1) simple key (single column), e.g. What is the \textbf{immigration} in \textbf{Salzburg}? 

(2) composite key (combination of 2 columns), e.g. What was the \textbf{immigration} in \textbf{Salzburg} in \textbf{2011}?

We generate data samples using a randomized procedure. For the first task we select a unique value for the key-column in each row and pick all other values uniformly at random from the respective domains. In order to learn successfully for the second task we generate unique rows that partially overlap in their composite keys, for instance: 

\begin{tabular}{lll}
\hline
Row2 & City & \textbf{Salzburg}\\
Row2 & Year & \textbf{2010}\\
Row3 & City & \textbf{Salzburg}\\
Row3 & Year & 2008\\
Row1 & City & Klagenfurt\\
Row1 & Year & \textbf{2010}\\
\hline
\end{tabular}\\[1ex]

These training examples explicitly require the model to attend to both columns that constitute the composite key. Otherwise, if a single column appears enough to uniquely identify the rows, the network ignores the second column of the composite key.

In order to avoid over-fitting when the network is memorizing the question template we provide 2 different question templates. At the data generation phase we select one of them uniformly at random for each training example. This aids the network in separating semantically important words (concepts from the table) from the connector words (\textit{in}, \textit{for}). This makes the model more flexible and robust in handling diverse question formulations which were not observed during the training phase.

\subsection{Evaluation}

In order to test the robustness of the trained model we create a test set with a single batch, where we take the generated data samples and change the test question by perturbing the template and paraphrasing the original question. We provide several scenarios that explore the ability of the model to recover the correct answer. The template-based questions are modified by

\textbf{omitting words}: one or more words are removed from the original user query;

\textbf{changing the position of words} in the query;

\textbf{querying a different column} that did not appear in the questions from the training data set;

\textbf{inadequate questions}, for which data required to answer this question are not present in the input table.

In this way we obtained a test set with 32 samples (8 samples for each of the 4 corruption types) with the questions phrased the way they never appeared in the pattern-generated training examples but are semantically meaningful and could occur in the real-world settings.

\begin{table}
\small
\centering
\begin{tabular}{|l|r|r|r|}
\hline \bf Task & \bf Test Error & \bf Training Set & \bf Epochs \\ \hline
Simple key & 0.5 & 5,949 & 29 \\
Composite key & 0.59 & 18,953 & 88 \\
\hline
\end{tabular}
\caption{\label{results} Evaluation results.}
\end{table}

The evaluation results are summarized in Table~\ref{results}. The error analysis showed that both models failed to provide the correct answer for the columns that never appeared in the questions of the training set. Also, the models output false answers in response to the questions for which the correct answer is not contained in the input table often with a high confidence, when relying on a single column from the composite key.

\section{Demonstration}

The aim of the demonstration is to showcase the power and limitations of the neural model trained to answer questions on semi-structured data. TableQA prototype is implemented as a Flask web application\footnote{Implementation based on \url{https://github.com/vinhkhuc/MemN2N-babi-python}} and is publicly available on our web-site (see \url{https://svakulenko.ai.wu.ac.at/tableqa}). 

The user interface (Figure~\ref{fig:ui}) allows to enter a custom question for a sample table provided (alternatively, use one of the questions from the test set held-out during the training phase). The attention weights are visualized by highlighting the corresponding cells in the input table, which provides an insight on the data patterns learned by the neural network. 

There is also an additional table below, which contains more details about the underlying prediction mechanism. It contains the triple-wise representation of the input table as consumed by the neural network and the attention weights for each of the three memory layers separately.

\section{Discussion of Limitations and Outlook}

The query disambiguation module is disjoint from the training module, which makes different types of errors more transparent. However, it uses an over-simplifying assumption that each word in the user query corresponds to a single word from the model vocabulary. A sequence-to-sequence model~\cite{DBLP:conf/emnlp/ChoMGBBSB14}, which is a common approach for language translation, in place of this simple heuristic could make query disambiguation more robust. Also, the pre-trained word embeddings can be integrated within a single neural network architecture to make the computation more efficient.

Our experiments showed that the design of the training examples is very important especially when trying to teach attention over the composite key for the second task. Also, the second task (composite key look-up) turned out to be much more difficult requiring more examples and time to train.

Since the model is trained exclusively on positive examples, i.e. correct question/answer pairs, it is incapable of handling inadequate user queries, i.e. questions that can not be answered using the provided input. This observation makes an obvious application in the real-world settings by demonstrating the need to train neural networks to identify and correctly handle such questions.

Another question type, not covered in the current evaluation, are ambiguous questions, which may relate to several cells at the same time. The model has to be able to identify such a situation and prompt the user to disambiguate the query or fall-back to the predefined behavior, e.g., output all relevant data or only the most recent ones.


The task for the future work remains in evaluating the model on the joint task including other bAbI datasets. Also, extending the model to work on the new data that was not available during the training phase, i.e. to propagate the learned weights to the OOV words, will make the approach applicable for the real-world data. This may involve changes in the network architecture, e.g. towards learning a hierarchical representation of the table structure that will create the necessary layers of abstraction beyond the individual values~\cite{DBLP:conf/ijcai/YinLLK16,DBLP:journals/corr/NeelakantanLAMA16}. The challenge, however, is to keep the network architecture general enough to perform well on other bAbI tasks at the same time to be able to answer questions of various kind and on different types of data (tables and text).

Finally, the web application can be further extended to accommodate user feedback and collect new annotations of question/answer pairs towards enriching the training dataset beyond the template-generated examples and improving the model.

\section{Conclusion}

In this paper we propose two new bAbI tasks for question answering from tables and provide an initial evaluation of the performance of the memory network architecture on them. These results can be used towards developing a natural-language interface that will support search in semi-structured data, such as Open Data tables.

The role of the demonstration is to provide an opportunity to interactively explore the performance and limitations of the trained model. It helps to understand which patterns the model has actually learned from the provided data samples. This tool will be useful for all who want to learn more about this family of models as well as for the researchers looking for directions to improve the neural network performance.

\section*{Acknowledgments}

This work was supported by the Austrian Research Promotion Agency (FFG) under the project CommuniData (grant no. 855407).

\bibliography{emnlp2017}
\bibliographystyle{emnlp_natbib}

\end{document}